\documentclass[graybox, envcountchap, authoryear, natbib]{svmult}

\usepackage{mathptmx}       
\usepackage{helvet}         
\usepackage{courier}        
\usepackage{makeidx}         
\usepackage{graphicx}        
\usepackage{multicol}        
\usepackage[bottom]{footmisc}

\begin{document}

\title*{Gas Accretion onto the Milky Way}
\author{Philipp Richter}
\institute{Philipp Richter 
\at 
Institut f\"ur Physik und Astronomie, Universit\"at Potsdam,
Karl-Liebknecht-Str.\,24/25, 14476 Golm, Germany
\email{prichter@astro.physik.uni-potsdam.de}
}
%
%
\maketitle

\abstract*{}
\abstract{
The Milky Way is surrounded by large amounts of gaseous matter that are
slowly being accreted over cosmic timescales to support star formation in
the disk. The corresponding gas-accretion rate represents a 
key parameter for the past, present, and future evolution of the
Milky Way. In this chapter, we discuss our current understanding of gas
accretion processes in the Galaxy by reviewing past and recent 
observational and theoretical studies. The first part of this review
deals with the spatial distribution of the different gas phases in
the Milky Way halo, the origin of the gas, and its total mass. 
The second part discusses the gas dynamics and the physical processes 
that regulate the gas flow from the outer Galactic halo to the
disk. From the most recent studies follows that the present-day gas 
accretion rate of the Milky Way is a few solar masses per year,
which is sufficient to maintain the Galaxy's star-formation 
rate at its current level.
}


\section{Introduction}
\label{sec:1}

We start this chapter on Galactic accretion with an introduction, in which
we first summarize early measurements of circumgalactic gas and the star-formation
history of the Milky Way from a historical perspective. We then 
shortly discuss the role of gas accretion processes in Milky-Way type galaxies
in a cosmological context and highlight their importance for galaxy evolution
in general. The introduction section ends with a proper definition 
of the gas-accretion rate, d$M_{\rm gas}$/d$t$, and an assessment of the physical
parameters that need to be constrained from observations and simulations  
to estimate d$M_{\rm gas}$/d$t$ for the Milky Way.

\subsection{Historical remarks}
\label{subsec:1}

The presence of gas above/below the Milky Way disk has been established
already more than 60 years ago, when absorption-line measurements
demonstrated the presence of gas clouds at high
galactic latitudes that exhibit relatively
high radial velocities (Adams 1949; M\"unch 1952; M\"unch \& Zirin 1961). 
In 1956, Lyman Spitzer argued that these structures, if located in the halo, 
must be surrounded by a hot, gaseous medium (he named this medium the {\it Galactic Corona}) 
that provides the necessary pressure-confinement of these clouds, otherwise they 
should disperse on relatively short timescales (Spitzer 1956). 

With the new receiver technologies and the resulting improved sensitivity of 
radio telescopes in the 1960s, high-velocity H\,{\sc i} 21 cm 
emission at high galactic latitudes was found by Muller et al.\,(1963),
Smith (1963), Dieter (1964), Blaauw \& Tolbert (1966), Hulsbosch \& Raimond (1966),
and Mathewson (1967), supporting the conclusions from the earlier 
absorption-line measurements. The observed distribution of radial velocities
of the 21 cm emission features lead to the definition of two classes
of Galactic halo clouds: as ``high-velocity clouds'' (HVCs) those halo structures
were labeled that have radial velocities, $|v_{\rm LSR}| > 100$ km\,s$^{-1}$,
while features with somewhat smaller radial velocities 
($|v_{\rm LSR}|\approx 30-100$ km\,s$^{-1}$) were given the name 
``intermediate-velocity clouds'' (IVCs). As will be discussed later, 
there also might exist a population of halo clouds with very low LSR velocities
(``low-velocity clouds'' (LVCs), similar to those in the disk.

Several scenarios for the origin of the 21 cm neutral halo clouds were discussed
by Jan Oort in 1966, among which the infall of intergalactic gas, the accretion
of gas from satellite galaxies, the condensation of neutral gas patches
from cooling coronal gas, and the ejection of gaseous material from the 
Milky Way disk were regarded as plausible scenarios (Oort 1966). Indeed,
as we will discuss in this review, these scenarios are still up-to-date.

The need for feeding the Milky Way disk with low-metallicity gas also
comes from early studies of the Galaxy's star-formation 
activity and stellar content. Already in the 1970s it was
realized that star-formation in the Milky Way would have come to
a halt early on, if the Galaxy was not fed with fresh material 
from outside. This is because the gas-consumption time scale
(even for a moderate star-formation rate of $\sim 1\,M_{\odot}$yr$^{-1}$)
is short compared to the age of the Milky Way (Larson 1972). 
Another argument for gas accretion comes from the observed metallicity 
distribution of low-mass stars in the solar neighborhood, which 
cannot be reproduced by closed-box models of the chemical evolution 
of the stellar disk. To match the observations, such models
{\it require} the continuous accretion of metal-poor gas  (van\,den\,Bergh 1962;
Chiappini et al.\,2001). These findings provide additional strong arguments that 
the Milky Way has accreted (and is still doing so) large amounts of
low-metallicity gas to continuously built up its stellar content 
as observed today.

In conclusion, the observed presence of large amounts
of gas above/below the disk, the past and present star-formation
rate of the Milky Way, and the metallicity distribution
of low-mass stars in the solar neighborhood demonstrate that gas accretion
represents an important process that has strong implications
for the past, present, and future evolution of our Galaxy.

\subsection{Cosmological context}
\label{subsec:2}

In the overall context of galaxy formation in the Universe,
gas accretion and feedback nowadays are regarded as the 
main processes that regulate the star-formation activity in 
galaxies. Many of the cosmological aspects of gas accretion in
galaxies will be discussed in detail in other chapters of this book. 
Still, for our discussion on gas accretion in the Milky Way, the 
main aspects need to be summarized here.

In the conventional sketch of galaxy formation, 
gas is falling into a dark matter (DM) halo and 
then is shock-heated to approximately the
halo virial temperature (a few $10^6$ K, typically), 
residing in quasi-hydrostatic equilibrium with the 
DM potential well (Rees \& Ostriker 1977). The gas 
then cools and sinks into the center of the potential 
where it is transformed into stars. This model is often
referred to as the `hot mode' of gas accretion. 
It has been argued that for smaller DM potential wells the 
infalling gas may reach the disk {\it directly} at much shorter 
timescales, without being shock-heated to the virial temperature
(`cold mode' of gas accretion; e.g., White \& Rees 1978; Kere{\v s} et al.\,2005). 
In the latter case, the star-formation rate of 
the central galaxy would be directly coupled to its gas-accretion
rate (White \& Frenk 1991). 
In these simple pictures, the dominating gas-accretion mode depends 
on the mass and the redshift of the galaxy
(e.g., Birnboim \& Dekel 2003; Kere{\v s} et al.\,2005),
where at low redshift the critical halo mass that separates 
the hot mode from the cold mode is $\sim 10^{12}\,M_{\odot}$
(van\,de\,Voort et al.\,2011). 
However, the underlying physics that describes the
large-scale flows of multi-phase gas from the outer to the inner
regions of a dynamically evolving galaxy is highly complicated
(e.g, Mo \& Miralda-Escude 1996; Maller \& Bullock 2004).
To understand these processes,
high-resolution hydrodynamical simulations with 
well-defined initial conditions (e.g, Bauermeister et al.\,2010;
Fumagalli et al.\,2011; van\,de\,Voort \& Schaye 2012;
Vogelsberger et al.\,2012; Shen et al.\,2013) are required. Therefore,
even the most advanced hydrodynamical simulations
do not provide tight constraints for the gas-accretion rate of 
{\it individual} galaxies without knowing their exact halo masses, 
their cosmological environment, and the initial circumgalactic gas 
distribution (Nuza et al.\,2014; hereafter referred to as N14).

Next to the feeding of the halos of Milky-Way type galaxies through 
intergalactic gas, galactic-fountain type processes (from supernova (SN)
feedback; Fraternali \& Binney 2008) and mergers with satellite galaxies 
(Di\,Teodoro \& Fraternali 2014) need to be considered.
The vast amounts of neutral and ionized 
gas carried by the Magellanic Stream underline the importance
of merger processes for the Milky Way's gas-accretion rate 
(D'Onghia \& Fox 2016; see Sect.\,2).
The preconditions under which such gas clouds are generated and falling 
toward the disk are different from those for clouds being accreted from
the intergalactic medium (IGM; e.g., Peek 2009) and 
thus they need to be explored separately by both observations and simulations.
Finally, it is important to keep in mind that the Milky Way is not an isolated
galaxy, but is embedded in the Local Group, being close to another
galaxy of similar mass, M31. The original distribution of gas that is entering 
the virial radius of the Milky Way from outside thus depends on the 
spatial distribution of satellite galaxies and the distribution
of intragroup gas in the cosmological filament that builds the
Local Group. In Fig.\,1 we sketch the local galaxy distribution in
the Local Group and the gas distribution around the Milky Way and M31 from
a hypothetical external vantage point. 

Turning back to the Milky Way, we know that since $z=1$ the Milky Way has produced 
$\sim 8\times 10^9 M_{\odot}$ of stars, while the current star-formation rate of the Milky 
Way is $\sim 0.7-2.3\,M_{\odot}$\,yr$^{-1}$ (Levine, Blitz \& Heiles 2006; Robitaille \& Whitney 2010;
Chomiuk \& Povich 2011; see also Peek 2009).
To relate this stellar mass and star-formation rate to the gas-accretion rate
it is important to remember that as much as 50 percent of the initial material from 
which a generation of stars is formed will be returned back to the ISM and will
be recycled in later stellar generations (Rana 1991).
This means that for one accreted mass unit of gas, {\it two} mass units 
of {\it evolved} stars will have emerged after several star-formation cycles.
For the Milky Way, this implies that is has accreted a gas mass of 
$\geq 4\times 10^9 M_{\odot}$ during the last 8 Gyr.


\begin{figure}[t!]
\includegraphics[width=11.7cm]{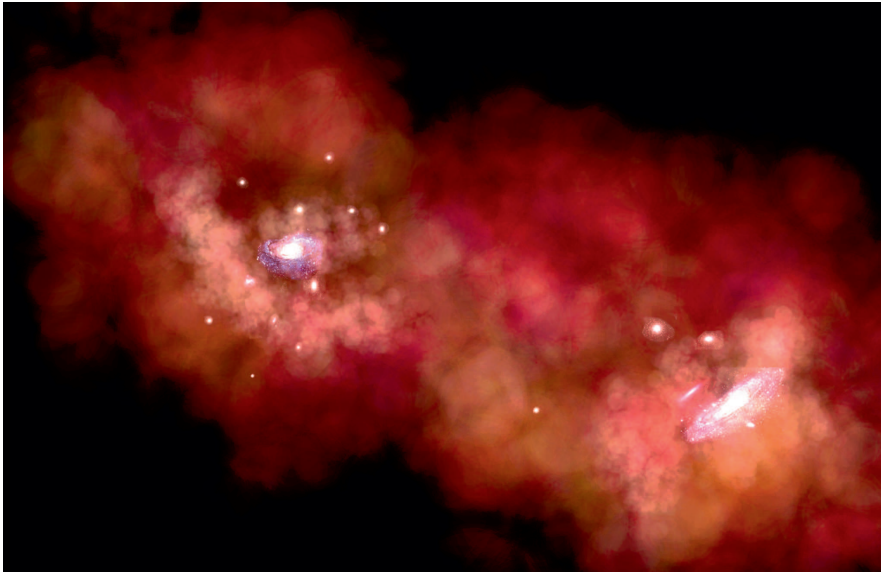}
\caption{
Sketch of the expected distribution of galaxies and multi-phase gas in the 
Local Group from an external vantage point. The Milky Way (left big galaxy) 
and M31 (right big galaxy) are surrounded by their populations of satellite
galaxies and by large amounts of multi-phase gas. Both galaxies are
interconnected by a gaseous bridge, which spatially falls together with
the Local Group barycenter (see, e.g., N14). Figure produced 
by the author for this review article.
}
\label{fig:1}       
\end{figure}


\subsection{Parameterization of gas accretion}
\label{subsec:3}

Before we start to discuss in detail the various observational
and theoretical aspects of gas-accretion processes in the 
Milky Way, we need a proper definition of the most important
parameters involved, in particular the accretion {\it rate}. 
From a cosmological perspective (e.g., in studies using cosmological
simulations), the growth and evolution of galaxies through cosmic times 
is governed by the gain of gas mass within their gravitational
sphere of influence
(i.e., within their virial radius, $R_{\rm vir}$) through
mergers and gas infall. In this context, one can simply define the gas 
accretion rate of a galaxy as the net mass inflow of gas through 
an imaginary sphere with radius $R_{\rm vir}$.
From a galaxy-evolution perspective, in contrast, the only relevant accretion
rate is that of the disk, where the infalling gas is being transformed into
stars, while the total amount of gas cycling within the galaxy's extended halo
is relatively unimportant. One of the most burning questions in gas accretion 
research therefore is, how much of the gas entering the virial radius
of a Milky-Way type galaxy actually makes it to the disk and what are the typical
timescales for this process?

In the following, we address these conceptional issues in two steps. First,  
we define the overall {\it current-day  gas-accretion rate} of the Milky
Way simply by relating the total mass of infalling gas, $M_{\rm gas}$, 
with its infall velocity, $v_{\rm infall}$, and its galactocentric distance,
$d$, so that


\begin{equation}
\frac{{\rm d}M_{\rm gas,halo}}{{\rm d}t}=\frac{M_{\rm gas}\,v_{\rm infall}}{d}.
\end{equation}


Because of the cloud's passage through the halo and the interaction with the
ambient hot coronal gas, only a fraction of this initial gas mass will end up 
in the disk to power star-formation therein. The {\it future} disk gas accretion 
rate thus can be defined as


\begin{equation}
\frac{{\rm d}M_{\rm gas,disk}}{{\rm d}t}=\eta \,\frac{{\rm d}M_{\rm gas,halo}}{{\rm d}t},
\end{equation}


where $\eta\leq 1$ represents the {\it fueling parameter} that modulates
the disk's gas accretion and star-formation rate at the time of impact.
The accretion time of each infalling gas cloud from its initial position seen
today to the disk is $t_{\rm acc}=d/\langle v_{\rm infall}\rangle$. Here, 
$\langle v_{\rm infall}\rangle$ represents the average 
infall velocity along the cloud's passage towards the disk.
In conclusion, it is the today's 3D distribution and space motion of gas 
around the Milky Way (parameterized by the current-day halo gas-accretion rate)
that governs the {\it future} star-formation activity in the Milky Way disk.

From equations (1) and (2) it becomes immediately clear, which parameters
need to be constrained by observations to get an insight into the gas accretion
rate of the Milky Way.


\begin{itemize}

\item
The {\bf total mass of halo gas} that is potentially flowing towards the
Milky Way disk needs to be constrained from observations. Such
observations need to take into account the huge span in physical
conditions such gas might have: from cold neutral gas (that can be observed
in H\,{\sc i} 21 cm emission) to million-degree shock-heated gas
(potentially visible in X-ray emission and absorption) all relevant
circumgalactic gas-phases need to be considered. Next to sensitivity
issues, disentangling distant halo gas from foreground disk gas is difficult
for material that has low radial velocities similar to those expected
for the rotating interstellar gas disk.

\item
Also the {\bf distance} of extraplanar/circumgalactic gas needs to be
determined by observational means. Reliable distance estimates for gas
complexes in the Milky Way halo are particularly challenging. The
so-called bracketing method requires the presence foreground and background
stars with known distances in the general direction of a
halo gas cloud to pinpoint a distance range for it
(Prata \& Wallerstein 1967). If high-velocity absorption of a given ion
is observed in the spectrum of a star with distance $d_{\rm star}$, 
it is clear that the absorbing gas is in front of the star at 
$d_{\rm gas}<d_{\rm star}$. The {\it significant} absence of absorption 
(for instance, taking into account the absorption strength predicted from the 
21 cm emission spectrum)
instead implies that the gas lies behind the star. This bracketing method is,
however, limited to large, spatially extended gas complexes and requires
a substantial observational effort. Other (indirect) distance estimates
for extraplanar/circumgalactic gas features (e.g., from modeling
the observed H$\alpha$ fluxes) are afflicted with systematic
uncertainties. 

\item
The {\bf infall velocity} of Milky Way halo gas cannot be observed directly,
but must be estimated from theoretical considerations in combination with
observational constraints. This is because the observed radial
velocities, $v_{\rm rad}$, of extraplanar/circumgalactic gas features do not
reflect their 3D space velocities. Also, the effect of galactic rotation adds
another velocity component that needs to be taken into account for the
interpretation of $v_{\rm rad}$. As infalling gas is expected to interact with
its ambient medium, the infall velocity is likely to depend on the overall
halo-gas properties and the position of the infalling gas cloud 
in the Milky Way potential well, i.e., on its distance to the disk.

\item
The {\bf fueling parameter $\eta$} can only be estimated from
full-fledged hydrodynamical models that realistically
describe the passage of neutral and ionized gas through the
hot coronal gas of the Milky Way. Such models need to take
into account the initial gas distribution of the infalling
and ambient medium and all relevant physical processes (e.g.,
instabilities, turbulent mixing layers, conductive interfaces,
and many others). In addition, possibly existing {\it outflowing}
gas components (e.g., from SN feedback) are likely to be
important.

\end{itemize}


In the following sections, we will discuss our understanding of these
individual parameters in detail, summarizing past and recent observational
and theoretical studies.


\section{The observed distribution of gas around the Milky Way}
\label{sec:2}

To pinpoint the 3D distribution of gas around the Milky Way and to
estimate its total mass, the different gas {\it phases} (in the
density-temperature phase-space) need to be considered. In the following,
we separate the Milky Way halo gas in three such phases:
(1) neutral/molecular gas, (2) warm ionized gas with temperatures
$T\leq 10^5$ K, and (3) hot ionized gas with $T> 10^5$ K. 
First, we discuss the angular distribution of these
phases, their galactocentric distances, and their total mass considering
recent observational results.
Then, we provide estimates on the gas-accretion rate of the Milky Way based on
these observations. Note that many of these aspects have also been discussed
in previous reviews (Wakker \& van\,Woerden 1997; Richter 2006; 
Putman, Peek \& Joung\,2012) and in the book on HVCs (van\,Woerden et al.\,2004).

\subsection{Neutral gas}
\label{subsec:1}

As discussed earlier, radio observations in the H\,{\sc i} 21 cm line have become the 
most powerful method  to study the distribution and internal structure of neutral 
halo clouds (Bajaja et al.\,1985; Hulsbosch \& Wakker 1988; Wakker \& van\,Woerden 1991; 
Hartmann \& Burton 1997; Morras et al.\,2000; Kalberla et al.\,2005;
McClure-Griffiths et al.\,2009; Winkel et al.\,2010). 
Absorption-line measurements against bright extragalactic background sources 
(e.g., quasars) provide additional information on the chemical composition of
the neutral gas and its connection to the other circumgalactic (ionized) gas phases.

In Fig.\,2 we show the sky distribution of the 21 cm IVCs (upper panel) and
HVCs (lower panel) plotted in galactic coordinates in an Aitoff-projection
centered on $l=180^{\circ}$.
Because of their different kinematics and intrinsic properties, it is
useful to discuss the properties of 21 cm IVCs and HVCs individually.

\subsubsection{IVCs}
\label{subsubsec:1}

The sky distribution of IVCs indicates that intermediate-velocity gas 
has predominantly negative radial velocities. 
The most prominent IVC features (see labels in Fig.\,2) are in the northern
sky the IV Arch and its low-latitude extension (LLIV Arch), IV Spur, the 
low-velocity part of Complex L, Complex K, and the low-velocity part of the 
Outer Arm. In the southern sky, there is 
the Anticenter (AC) shell and the Pegasus-Pisces (PP) Arch
(see Kuntz \& Danly 1996; Wakker 2001).
The total sky covering fraction of neutral IVC gas is $f_{\rm c}\approx 0.30$ 
for column densities $N$(H\,{\sc i}$)\geq 10^{19}$ cm$^{-2}$ 
and deviation velocities $|v_{\rm dev}|=30-90$ km\,s$^{-1}$ 
(Wakker 2004; see caption of Fig.\,2). Many of the IVC features are spatially and 
kinematically connected with 21 cm disk gas.
The typical H\,{\sc i} column densities in IVCs in the 21 cm surveys lie in the 
range $N$(H\,{\sc i}$)=10^{19}-10^{20}$ cm$^{-2}$, but some cores
in the IV Arch and the Outer Arm exhibit column densities above
$10^{20}$ cm$^{-2}$ (Fig.\,2).
At positive radial velocities, there is the low-velocity extension of 
HVC Complex WA in the north and Complex gp in the south. The positive
velocity IVCs are much smaller in angular size and column density and
they exhibit a very complex internal structure.

Direct distance measurements (mostly using the bracketing technique) place the
IVCs relatively close to the Milky Way disk with $z$ heights $<2.5$ kpc,
typically. The most massive
IVC complexes, for example, have distances of $0.8-1.8$ kpc (IV Arch),
$\sim 0.9$ kpc (LLIV Arch), and $0.3-2.1$ kpc (IV Spur; see IVC compilations by 
Wakker 2001, 2004 and references therein).
Most of the IVCs thus belong to the disk halo interface, which possibly
represents a crucial component for gas accretion processes in the Galaxy, as will be
discussed below. For the most prominent, large-scale IVC complexes, the
21 cm data indicate individual H\,{\sc i} masses on the order of
$1-8\times 10^5 M_{\odot}$ (e.g., IV Arch, LLIV Arch, IV Spur;
Wakker 2001). The total neutral gas mass of the large-scale IVCs thus can estimated
to be $\sim 10^6 M_{\odot}$.

Absorption-line measurements indicate that 
the metallicity of most of the large northern IVCs is relatively high with 
typical values between $0.5$ and $1.0$ solar (e.g., Wakker 2001; Richter et 
al.\,2001a, 2001b). For the Outer Arm, Tripp \& Song (2012) derive a lower metallicity 
of $0.2-0.5$ solar, suggesting that this gas (albeit being close to the disk) 
might have an extragalactic origin,
such as many HVCs (see below).
Another interesting feature is the core IV\,21, which has a metallicity of just
$0.4$ solar (Hernandez et al.\,2013) at a $z$ height of $\sim 300$ pc 
above the disk, thus also pointing toward an extragalactic origin.
Dust is also present in IVCs, as is evident from the observed depletion 
patterns of heavy elements in intermediate-velocity gas (Richter et al.\,2001a, 2001b;
Wakker 2001; see also Savage \& Sembach 1996)
and from the observed excess of extra-planar infrared 
emission in the direction of IVCs (Desert et al.\,1988, 1990; Weiss et al.\,1999).
Also molecules (H$_2$, CO) have been detected
in several IVCs (Richter et al.\,2001b, 2003; Wakker 2006; Gillmon et al.\,2006;
Hernandez et al.\,2013; R\"ohser et al.\,2016), 
but the molecular gas fraction is very small, so that the molecular phase does
not contribute significantly to the overall IVC masses. Yet, the presence of 
molecular gas implies the presence of substantial small-scale structure
in the gas down to AU scales (Richter, Sembach \& Howk 2003).
Small-scale structure is also evident from high-resolution 21 cm data, which
show, next to the coherent large-scale 21 cm IVC complexes, a population of several
thousand compact H\,{\sc i} clumps at $z=1-2$ kpc (e.g., Lockman 2002; 
Kalberla \& Kerp 2009; Saul et al.\,2012). These clumps, that have
very small masses of only $10^1-10^4 M_{\odot}$, may represent 
cloudlets that have condensed out of the ambient multi-phase medium. They
are raining down to the disk, thus fueling it. Because of their small
cross section for absorption spectroscopy, the metal and dust content 
of these clumps remains unknown so far.

In view of the measured overall chemical composition of the 
large IVCs and their location in the disk-halo interface, the favored scenario 
for the origin of near solar-metallicity IVCs is, that these structures represent
the back-flow of cooled gas from the galactic fountain process 
(Shapiro \& Field 1976; Houck \& Bregman 1990); i.e.,
they originate from metal-enriched gas that has been ejected from the disk by 
supernova explosions that is now cycling back to the disk due to gravitational forces.


\begin{figure}[t!]
\centering
\includegraphics[width=10.7cm]{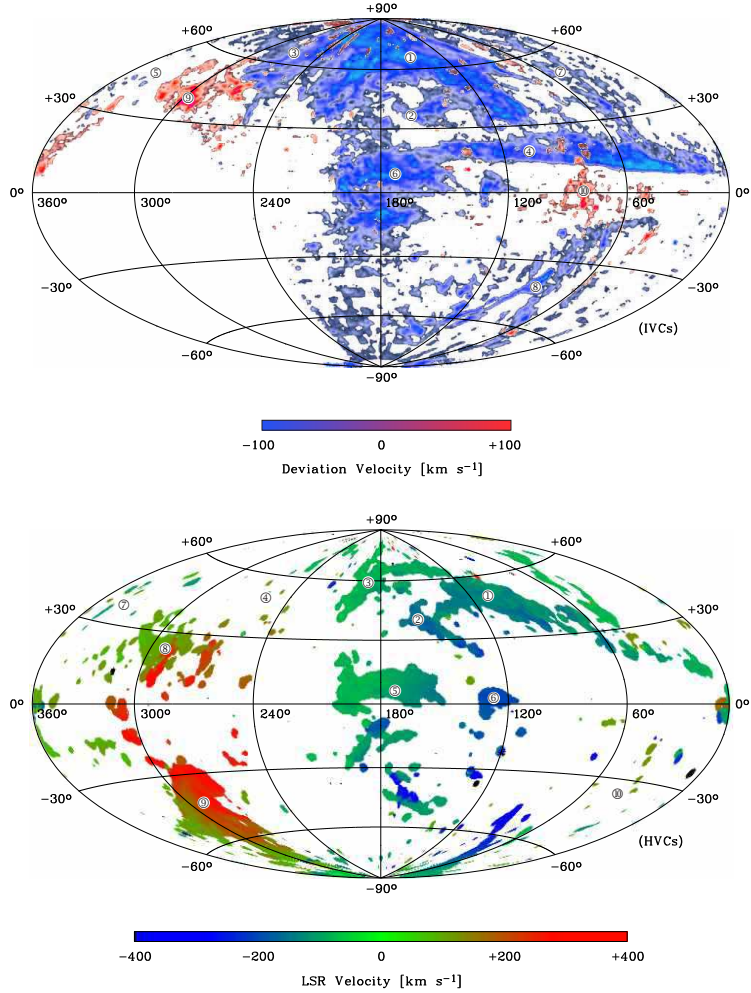}
\caption{
All-sky maps of 21 cm emission for intermediate-velocity gas
(upper panel) and high-velocity gas (lower panel) in 
an Aitoff projection centered on $l=180^{\circ}$. 
The maps show the sky distribution of neutral gas in the Galactic halo.
The maps have been generated from different data sets described in Wakker (2004),
Kalberla et al.\,(2005), and by Tobias Westmeier (priv.\,comm.).
For the IVCs we show (color-coded) the deviation velocity of the gas 
from a simple model of Galactic rotation (see Wakker 2004) in the 
range $|v_{\rm dev}|=30-90$ km\,s$^{-1}$. For the HVCs, we display the 
color-coded LSR velocity ($|v_{\rm LSR}|=100-500$ km\,s$^{-1}$).
Individual neutral IVC and HVC complexes (see Wakker 2001, 2004) are labeled 
with numbers. For IVCs:
(1) IV Arch, (2) LLIV Arch, (3) IV Spur, (4) Outer Arm, (5) Complex L, 
(6) AC Shell, (7) Complex K, (8) PP Arch, (9) IV-WA, (10) Complex gp.
For HVCs: (1) Complex C, (2) Complex A, (3) Complex M, (4) Complex WA, 
(5) AC Cloud, (6) Complex H, (7) Complex L, (8) Leading Arm of MS, 
(9) Magellanic Stream, (10) Complex GCN.
}
\label{fig:2}       
\end{figure}


\begin{figure}[t!]
\centering
\includegraphics[width=10.7cm]{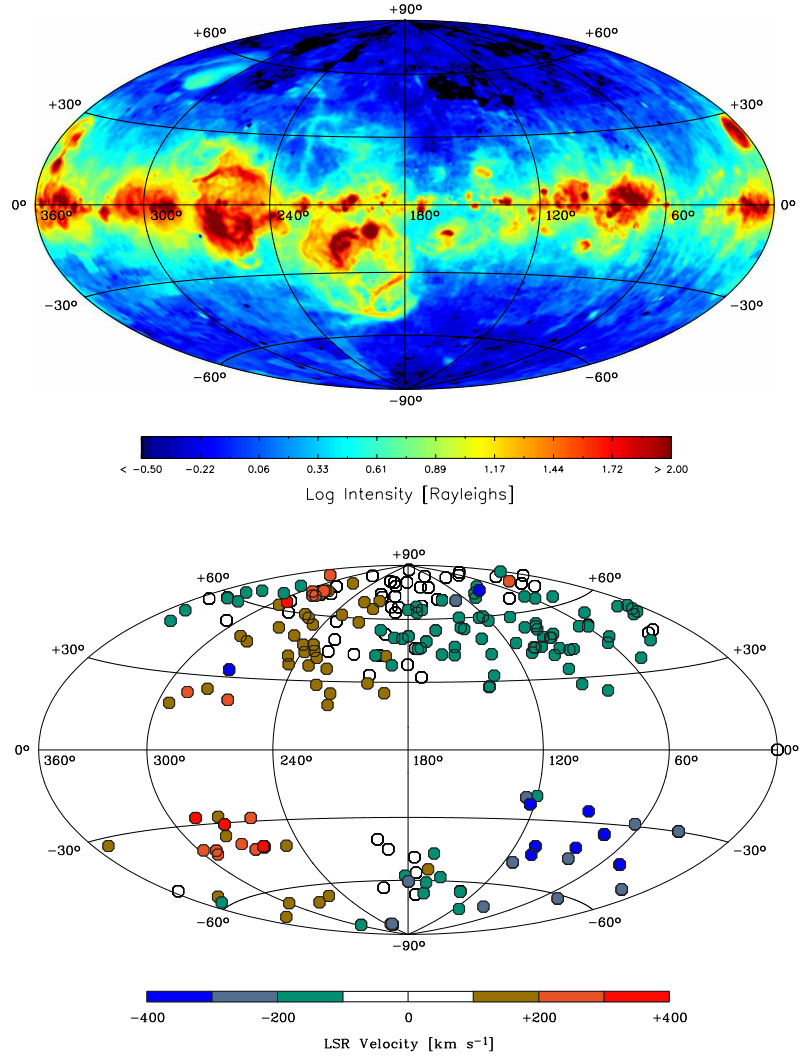}
\caption{
All-sky maps of H$\alpha$ emission in the Milky Way (upper panel) 
and high-velocity UV absorption of Si\,{\sc iii} absorption (lower panel)
in an Aitoff projection centered on $l=180^{\circ}$
The maps show the sky distribution of diffuse ionized gas (DIG) in the lower and upper 
Galactic halo, respectively.
The H$\alpha$ map (kindly provided by Matt Haffner) has been compiled from data
of the Wisconsin H$\alpha$ mapper (WHAM; Haffner et al.\,2003, 2016).
It shows emission from relatively dense gas that predominantly resides in 
the inner halo and disk-halo interface of the Milky Way (DIG layer;
Reynolds 1991).
The Si\,{\sc iii} data stem from the high-velocity 
($|v_{\rm LSR}|=100-500$ km\,s$^{-1}$) UV absorption survey 
from Richter et al.\,(2017; hereafter referred to as R17) 
using 265 HST/COS spectra of extragalactic 
background sources. Diffuse ionized gas has a substantially larger
sky covering fraction ($\sim 2-3$ times higher) than the neutral gas
(Fig.\,2).
}
\label{fig:3}       
\end{figure}


\subsubsection{HVCs}
\label{subsubsec:2}

The sky distribution of the Milky Way HVCs is far more complex than
that of the IVCs (Fig.\,2, lower panel); HVCs span a huge LSR velocity range of 
$\sim 700$ km\,s$^{-1}$. 

Among the most prominent Milky HVCs is Complex C, which covers
$\sim 1500$ square-degree on the northern sky in 21 cm emission, which
is about 4 percent of the entire sky (Wakker 2004). Complex C is a cloud
that presumably is being accreted from the IGM or from a satellite galaxy 
(e.g., Sembach et al.\,2004). 
Another prominent 21 cm HVC is the Magellanic Stream (MS) in the south.
The MS also covers an area of $\sim 1500$ square-degree in 21 cm, but it spreads over 
the entire southern sky, forming a coherent stream of neutral gas (D'Onghia \& Fox 2016).
The MS represents a tidal feature expelled from the Magellanic Clouds as they approach
the Milky Way halo (e.g., Gardiner \& Noguchi 1996; Connors et al.\,2006; 
Besla et al.\,2010, 2012; Nidever et al.\,2010; Diaz \& Bekki 2011, 2012). 
Over its entire body the MS spans a distance range 
of $d=50-100$ kpc (or even further) from the Galactic disk, 
suggesting that it extends over several 
hundred kpc in the outer halo of the Milky Way (Putman et al.\,1998, 2003; Stanimirovic et 
al.\,2002, 2008; Br\"uns et al.\,2005).
Other prominent Galactic HVCs are Complex A, Complex H, the Anti-Center Cloud,
and Complexes WA$-$WE. Their positions are indicated in Fig.\,2. 
The individual properties of all HVC complexes are discussed in detail in
Wakker (2001). 

The entire HVC population of the Milky Way shown in Fig.\,2 has a total 
sky covering fraction in 21 cm of $f_{\rm c}\approx 0.35$ for neutral gas 
column densities $N$(H\,{\sc i}$)\geq 7\times10^{17}$ cm$^{-2}$ 
(Murphy, Lockman \& Savage 1995; Wakker 2004 and references therein). The covering fraction reduces to 
$f_{\rm c}\approx 0.15$ for larger column densities 
$N$(H\,{\sc i}$)\geq 2\times10^{18}$ cm$^{-2}$. The H\,{\sc i} column 
densities in HVCs follow a well-defined column-density distribution function 
of the form $f(N_{\rm HI})\propto N_{\rm HI}^{-\beta}$ with $\beta=1.42$
for log $N$(H\,{\sc i}$)\geq 18$ (Lockman et al.\,2002). With the 
exception of the MS, all HVCs for which direct distance information 
from the bracketing method is available, are located within 20 kpc. For
instance,  the Complexes A \& C have distances $d\approx 10$ kpc (van\,Woerden et al.\,1999;
Wakker et al.\,2007; Smoker et al.\,2011; Thom et al.\,2008), the Cohen Stream and Complex GCP are
at $d=5-15$ kpc (Wakker et al.\,2008), and the HVC towards the LMC has 
$d\approx 9$ kpc (Richter et al.\,2015). For these structures in the inner Milky Way halo 
the distances estimated indirectly from the measured H$\alpha$ fluxes agree well 
with the values derived from the bracketing method (e.g., Bland-Hawthorn
\& Putman 2001; Putman et al.\,2003; Tufte et al.\,2002).
Combining the 21 cm emission in Milky Way HVCs and in M31 halo clouds, 
Richter (2012) predicts that the H\,{\sc i} covering fraction in HVCs around
Milky Way-type galaxies declines exponentially with galactocentric
distance with $f_{\rm c}<0.01$ for $d>70$ kpc. From deep 21 cm observations 
of the M31 halo (Westmeier et al.\,2007) further follows that also the so-called 
compact high-velocity clouds (CHVCs), isolated high-velocity 21 cm gas clumps
with very small angular sizes of $<2\deg$ (Braun \& Burton 1999; de\,Heij,
Braun \& Burton 2002), are located at $d<50$ kpc, disproving a previous 
scenario in which CHVCs are regarded as gas-filled dark matter halos residing
in the Local Group (Blitz et al.\,1999; Braun \& Burton 1999).

From the 21 cm data and the available distance information it follows that the 
total neutral HVC gas mass is $M_{\rm HI,HVC}\approx 2.5\times 10^8\,M_{\odot}$
(Wakker 2004; Br\"uns et al.\,2005). The MS contributes with more 
than $60$ per cent to this mass ($M_{\rm HI,MS}\approx 1.6\times 10^8\,M_{\odot}$
for $d=55$ kpc), while the other HVCs have substantially smaller masses (e.g., 
$M_{\rm HI,Complex C}\approx 5\times 10^6\,M_{\odot}$; Wakker 2004). Since the 
two Magellanic Clouds are located at only $\sim 80$ kpc distance, their 
interstellar gas content adds to the gas mass being accreted by the 
Milky Way. Therefore, if we add the neutral gas mass of the Magellanic Clouds 
and their gaseous interconnection, the Magellanic Bridge, the 
total H\,{\sc i} budget in the Milky Way halo sums up to a value of
$M_{\rm HI,MWhalo}\approx 1.3\times 10^9\,M_{\odot}$, which is $\sim 20$ percent 
of the neutral gas mass in the Milky Way disk 
($M_{\rm HI,MWdisk}\approx 7\times 10^9\,M_{\odot}$; Ferriere 2001).

Other tracers of predominantly neutral gas in the Milky Way halo are absorption lines 
of neutral and low ions, such as O\,{\sc i}, N\,{\sc i}, Ar\,{\sc i}, S\,{\sc ii} 
with transitions in the UV, and Ca\,{\sc ii} and Na\,{\sc i} in the optical regime.
From a survey of high-velocity Ca\,{\sc ii}/Na\,{\sc i} absorption in the Milky Way halo
against several hundred extragalactic background sources, Ben Bekhti et al.\,(2008, 2012)
derive a covering fraction of $f_{\rm c}\approx 0.50$ for log $N$(Ca\,{\sc ii}$)>11.4$,
which is $\sim 2$ times higher than the 21 cm covering fraction. These results further
indicate the widespread presence of cold, neutral gas structures away from the 
large 21 cm complexes.  
Such structures possibly are too small to be seen in all-sky 21 cm survey because of the
limited angular resolution of these surveys in combination with beam-smearing effects.

While most of the IVCs have near-solar metallicities, the metal abundance in many HVCs
is substantially lower (by a factor $5-10$, typically). 
For Complex C, the (mean) metallicity has been constrained to $0.15$ solar 
(Wakker et al.\,1999; Richter et al.\,2001a; Tripp et al.\,2003; Collins, Shull \& Giroux 2003;
Sembach et al.\,2004).
The main body of 
the Magellanic Stream also has a metallicity of only $0.1$ solar (Fox et al.\,2010, 2013),
but the MS contains a filament that is more metal rich ($0.3$ solar; Richter
et al.\,2013; Gibson et al.\,2001). Similarly, Complex A most likely has metallicity 
of only $\sim 0.1$ solar (Wakker 2004).
Such low metallicities are in line with 
the idea that HVCs represent gas infalling from the pre-enriched intergalactic medium 
(or intragroup gas), but the clouds may also trace material stripped from 
satellite dwarf galaxies as they are being accreted by the Galaxy. Moreover, 
it cannot be ruled out that some of the gas has been part of the Milky Way a 
long time ago, then was ejected (at relatively low metallicity) by a former Galactic 
outflow or wind, and now is raining back towards the disk ("intergalactic fountain").
An example for this latter scenario is the Smith Cloud (also called Complex GCP),
which has a metallicity of $0.5$ solar and is believed to originate in the
outer Galactic disk (Lockman et al.\,2008; Hill, Haffner \& Reynolds 2009; Fox et al.\,2016).

There appears to be only little dust in neutral and ionized HVCs 
(e.g., Wakker \& Boulanger 1986; Bates et al.\,1988; Tripp et al.\,2003; 
Richter et al.\,2001, 2009; Williams et al.\,2012).
The few tentative detections of far-IR emission in some HVCs 
(e.g, Miville-Desch{\^e}nes et al.\,2005; Peek et al.\,2009; 
Planck Collaboration 2011) remain inconclusive 
with respect to their dust abundance. Diffuse molecular gas is present only in
high-column density regions of the Magellanic Stream 
(Sembach et al.\,2001; Richter et al.\,2001c, 2013), the Magellanic Bridge
(Lehner 2002; Murray et al.\,2015), and in a dense clump of an HVC 
projected onto the LMC (Richter et al.\,1999). However, the molecular component 
is not of importance for the total mass of HVCs.

Finally, it is worth noting that some of the above mentioned neutral halo 
clouds (e.g., Complex L) exhibit a radial
velocity range that extends from values below $100$ km\,s$^{-1}$ to values above 
this threshold, i.e., these complexes can be regarded as {\it both} IVCs and HVCs, 
although they each represent a single, kinematically coherent structure. 
The question arises, whether the separation of IVCs
and HVCs as different halo-cloud populations is justified, or whether they just represent
the same population of objects with just different radial velocities. As discussed above,
distance measurements place the IVCs within 2.5 kpc of the Milky Way disk, while 
most of the HVCs are located much further away. This, together with the on average
higher metallicity of IVCs compared to HVCs, indeed indicates that low-velocity 
halo clouds (with LSR velocities {\it typically} $<100$ km\,s$^{-1}$) predominantly reside
in the disk-halo interface, while high-velocity halo clouds (with LSR velocities 
{\it typically} $\geq 100$ km\,s$^{-1}$) predominantly trace gas at $d=2-100$ kpc.

\subsection{Warm ionized gas}
\label{subsec:2}

Warm ionized gas in the Milky Way halo is even more widespread than the neutral
gas traced by 21 cm emission. We here define warm-ionized gas as gas that is 
predominantly ionized (i.e., with small neutral gas fractions) and has
a temperature $<10^5$ K. Circumgalactic gas at such temperatures is expected 
to be photoionized by the combined ionizing radiation from stars in the Milky Way disk and
the ambient extragalactic UV background at $z=0$ (see model by Fox et al.\,2005).
Warm ionized gas in the Milky Way halo can be detected either in emission
in recombination lines such as H$\alpha$ or in high-velocity 
UV absorption of so-called intermediate ions that have lower ionization boundaries
in the range $15-40$ eV (e.g., C\,{\sc iii}, N\,{\sc iii}; Si\,{\sc iii}, 
Si\,{\sc iv}, Fe\,{\sc iii}; Morton 2003; R17).

In Fig.\,3, upper panel, we show the sky distribution of H$\alpha$ emission in
the range $0.3-100$ Rayleigh, based on data obtained from the Wisconsin H$\alpha$ Mapper
(WHAM; Haffner et al.\,2003, 2016). As can be seen, H$\alpha$ emission is widespread at latitudes
$b<30^{\circ}$, situated in distinct coherent spatial structures such as lobes
and arches. These features reflect the complex motions of diffuse 
ionized gas (DIG) in the disk-halo interface (DHI) that is believed to
be shaped by the on-going star-formation in the Milky Way disk. 
Based on the derived gas densities and volume filling factors 
(Reynolds et al.\,2012; Haffner et al.\,2003), the total mass of ionized gas in the DHI
of the Milky Way can be estimated to be $\sim 10^8 M_{\odot}$,
in line with estimates for extra-planar ionized gas in other
low-redshift disk galaxies (e.g., NGC\,891; Dettmar 1990).


\begin{figure}[t!]
\centering
\includegraphics[width=12cm]{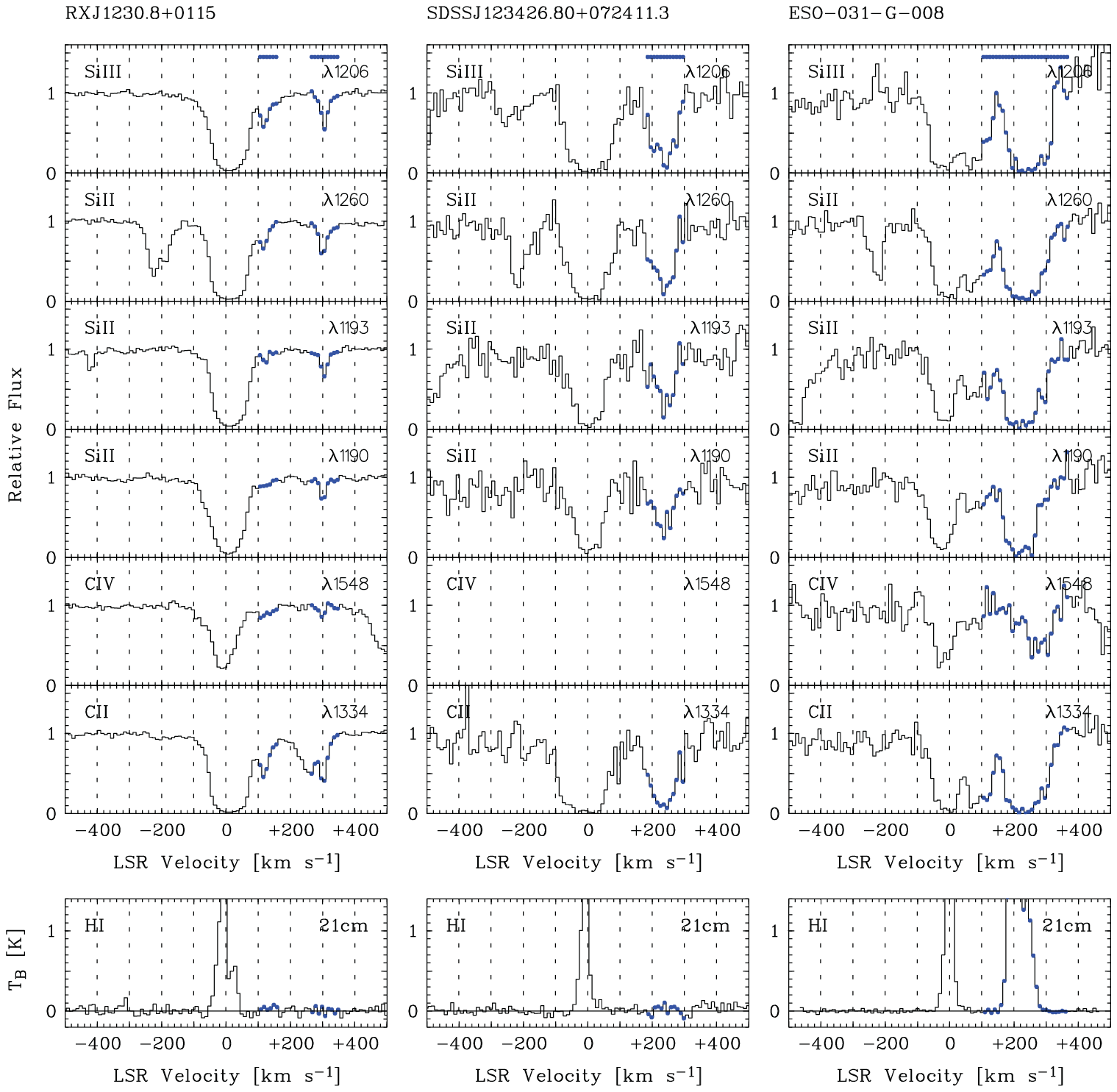}
\caption{
Velocity profiles of high-velocity UV absorbers in the Galactic halo along three 
lines of sight (from R17), based on HST/COS data.
Absorption profiles of various
transitions from Si\,{\sc iii}, Si\,{\sc ii}, C\,{\sc iv}, and C\,{\sc ii} 
are shown, where the high-velocity absorption at $|v_{\rm LSR}|\geq 100$ km\,s$^{-1}$ 
is indicated in blue. We also show the H\,{\sc i} 21 cm emission profiles 
(from GASS/EBHIS data) for the same
sightlines in the lowest panels. The individual high-velocity absorption 
features towards RXJ1230.8+0115 and SDSSJ123426.80+072411.3 
trace gas streams in the halo that are predominantly ionized 
(without H\,{\sc i} 21 cm counterpart), while the high-velocity
absorption towards ESO$-$031$-$G$-$008 traces neutral {\it and}
ionized halo gas related to the MS (UV absorption
plus H\,{\sc i} 21 cm emission).
}
\label{fig:4}       
\end{figure}


Many of the 21 cm HVCs at much larger distances from the disk are also detected 
in H$\alpha$ emission (e.g., Weiner \& Williams 1996; Tufte, Reynolds \& Haffner 1998;
Bland-Hawthorn et al.\,1998), 
indicating that the neutral gas clouds are surrounded by envelopes 
of ionized gas, whose masses are comparable with or even larger than
the neutral gas body (e.g., Fox et al.\,2004).
Also the Magellanic Stream at $d=50-100$ kpc is detected in H$\alpha$ 
(e.g., Putman et al.\,2003; Fox et al.\,2014), proving that the MS
is surrounded by substantial amounts of warm H\,{\sc ii}. From the 
models of Bland-Hawthorn et al.\,(2007) it follows that the mass-weighted
H\,{\sc ii} column density in the MS is $>10^{20}$ cm$^{-2}$, thus larger than
the mass-weighted H\,{\sc i} column density.
A similar conclusion was drawn by Fox et al.\,(2014), who determine
an ionized-to-neutral hydrogen mass ratio of $\sim 3$ based on the
absorption strength of intermediate and high ions associated with
the 21 cm body of the Stream.

The most sensitive ions to trace the warm ionized gas in the halo are C\,{\sc iii}
and Si\,{\sc iii} with strong transitions in the UV at $977.02$ \AA\,(C\,{\sc iii}) and 
$1206.50$ \AA\,(Si\,{\sc iii}; see Richter et al.\,2016). 
In their recent legacy survey of high-velocity 
UV absorption in the Milky Way halo, R17 found that 
warm ionized halo gas, as traced by Si\,{\sc iii} at velocities 
$|v_{\rm LSR}|\geq 100$ km\,s$^{-1}$ and column densities 
log $N$(Si\,{\sc iii}$)\geq 12.1$, has a covering fraction as high as 
$f_{\rm c}=0.74$, 
confirming earlier results based on much smaller samples (Collins, Shull \& Giroux 2009; 
Shull et al.\,2009; Lehner et al.\,2012; Herenz et al.\,2013). This 
covering fraction is more than twice the value obtained for the neutral 
HVCs from the 21 cm observations (see above). In the lower panel of Fig.\,3 
we show the sky distribution of high-velocity Si\,{\sc iii} absorption, 
which can be directly compared with 
the 21 cm HVC map (Fig.\,2, lower panel). Like the H$\alpha$ emission, high-velocity 
Si\,{\sc iii} absorption is often associated in radial velocity with the 
21 cm HVC features, even if located several degrees away from the 21cm contours.
This further implies that HVCs represent coherent multi-phase gas streams (with
a neutral gas body surrounded by an ionized gas layer) that move 
through the Milky Way halo (Lehner et al.\,2012).
Figs.\,2 and 3 also indicate that there are several regions in the high-velocity sky
that exhibit pronounced Si\,{\sc iii} absorption but do not show significant
large-scale H\,{\sc i}. 
In contrast to the 21 cm HVCs and their ionized envelopes, gas in these regions
can be regarded as {\it coherent ionized gas streams} in which patchy condensations
of cooler, neutral gas clumps are embedded.
Particularly interesting are the regions 
$l>200^{\circ}, b>0^{\circ}$ and $l<120^{\circ}, b<0^{\circ}$,
which form a velocity dipole on the sky in UV absorption (Fig.\,2; R17;
Collins, Shull \& Giroux 2003) in a direction
that forms the major axis of the Local Group cosmological filament (N14). 
The observed kinematically distinct absorption features at high positive and
high negative radial velocities possibly indicate that the Milky Way is ramming
into ionized intragroup gas because it follows the general flow of galaxies in the 
direction of the Local Group barycenter (Peebles et al.\,2001, 2011; Whiting 2014), 
while it is moving away
from Local Group gas in the opposite direction along the filament (R17).
Therefore, the Milky Way's accretion of warm ionized gas might be strongly influenced
by the local galaxy environment and cosmological structure formation in the Local
Group (N14). This important aspect will be further discussed
in Sect.\,3.2.

The ionized gas components that are associated with the 21 cm features obviously have 
the same distances as the neutral halo clouds. They also have comparable metallicities,
if the gas has not yet been mixed with the ambient hot coronal gas. This implies that the majority of
the diffuse ionized halo clouds that are {\it not} associated with the MS are located at 
$d<20$ kpc, while the ionized envelope of the Stream is at $d=50-100$ kpc. The exact
angular extent of the Stream's ionized gas component is unknown, but it may well
cover $30-50$ percent of the entire sky (R17; Fox et al.\,2014).
From their survey of high-velocity UV absorption towards Galactic halo stars with
known distances ($d<15$ kpc) and extragalactic background sources 
Lehner \& Howk (2011) and Lehner et al.\,(2012) find that the sky-covering 
fraction of high-velocity UV absorption increases only marginally 
from the halo-star sample to the QSO sample, if velocities $|v_{\rm LSR}|\leq 170$
km\,s$^{-1}$ are considered. This indicates that HVCs in this velocity range are
predominantly located at $d<15$ kpc. In contrast, HVCs with absolute LSR velocities 
larger than $170$ km\,s$^{-1}$ (e.g., the MS) are only seen against extragalactic 
background sources, demonstrating that the gas is located at $d>15$ kpc, and being in
line with the distance constraints for the neutral gas (see above).
The halo-star sample of Lehner \& Howk (2011) and Lehner et al.\,(2012) 
covers only a limited fraction of the sky, however, so that 
the possible presence of more distant ionized gas structures even at low velocities 
(in particular in the directions of a possible Local Group filament) cannot be 
ruled out with these data.

The total mass of diffuse ionized high-velocity gas in the Galactic halo 
is dominated by the extended envelope of the MS (Fox et a.\,2014; R17).
Assuming $d=55$ kpc and calculating the amount of H\,{\sc ii} from the
observed ion abundances in combination with an ionization model, both studies
obtain a gas mass of the ionized component of the MS of 
$M_{\rm MS}\approx 1-3 \times 10^9\,M_{\odot}$. 
This mass would be substantially higher, if some of the gas from the MS
was located at larger distances. For instance, if the distance of MS would lie in the 
range $d=100-150$ kpc (see Besla, et al.\,2012; Jin \& Lynden-Bell 2008; 
Bland-Hawthorn et al.\,2013), the mass of the ionized component of the MS would be as large 
as $\sim 3-7 \times 10^9\,M_{\odot}$, thus very close to the total ISM gas mass 
in the Galactic disk (Ferriere 2001).
The contribution of high-velocity absorbers at $d<20$ kpc to the ionized 
gas mass in the halo is small instead; their gas mass sums up to a total value 
of no more than $M_{\rm HVCs,d<20 kpc}= 2 \times 10^7\,M_{\odot}$ 
(R17). This value still is comparable to or even higher than
the mass of the neutral gas in the same distance range.

\subsection{Hot ionized gas}
\label{subsec:3}

Ever since the prediction of Lyman Spitzer in 1956 on the existence
of a Galactic Corona (see Sect.\,1), the search for a low-density,
high-temperature ($T>10^5$ K) gaseous medium that surrounds the Milky Way has 
been of high priority for astrophysicists, as the Corona links the 
observed properties of the Galaxy to cosmological structure formation 
(see, e.g., Oort 1966).
From more recent theoretical work (e.g., Maller \& Bullock 2004), it is indeed 
expected that all MW-type galaxies are surrounded by massive, hot gaseous halos 
of typical mass of $10^{11} M_{\odot}$ and temperature $T\sim 10^6$ K (the virial
temperature of the galaxy's DM halo).
If some fraction of the gas was able to cool, it would sink towards the disk, 
feeding the galaxy with fuel for future star formation. Therefore, hot coronal gas 
may serve as a huge baryon reservoir from which MW-type galaxies gain their 
gas. In addition, the hot Milky Way halo might be further fed with gas from 
a possible large-scale outflow from the Galactic center region
(Fox et al.\,2015; Lehner et al.\,2012; Zech et al.\,2008; Bland-Hawthorn \& Cohen 2003;
Su, Slayter \& Finkbeiner 2010).

Despite the obvious importance of the hot, ionized circumgalactic gas phase for
galaxy evolution, our knowledge about the properties and spatial
extent of hot coronal gas in the Milky Way still is very limited. 
This is because it is very difficult to detect such coronal gas that is expected to have
very low densities ($n_{\rm H}<10^{-3}$ cm$^{-3}$), in particular in
the outer regions of the halo. 
Observational evidence for the existence of a hot Milky Way Corona comes from
observations in the X-ray regime, where the gas can be observed in 
emission or in absorption against extragalactic X-ray
point sources. Using ROSAT data, Kerp et al.\,(1999) systematically searched
for X-ray emission spatially associated with neutral HVCs and reported
several positive detections, e.g., in the direction of Complex GCN, Complex C,
and Complex D. In Fig.\,5, we show as an example the ROSAT emission map of 
hot halo gas in the direction of Complex GCN from the study
of Kerp et al.
Other observations of the soft X-ray background support the 
existence of hot coronal gas in the Milky Way (e.g., Kuntz \& Snowden 2000).
X-ray absorption of O\,{\sc vii} and O\,{\sc viii} in the Galactic halo has been reported 
by several groups (Wang et al.\,2005; Fang et al.\,2002, 2003, 2006; Mathur et al.\,2003; 
Bregman 2007; McKernan, Yaqoob \& Reynolds 2004;
Williams et al.\,2005; Gupta et al.\,2012; see also Miller, Kluck \& Bregman 2016),
but the interpretation of these low-resolution spectra is afflicted
with systematic uncertainties (see Richter, Paerels \& Kaastra 2008).
Also pulsar dispersion measures have been used to constrain the
properties of hot coronal gas in the Milky Way halo (e.g., Gaensler et al.\,2008).
All these observations are biased towards the regions
with the highest gas densities in the Corona, however, so that the bulk of the hot gas 
detected in this manner presumably resides in the inner halo at $d<20$ kpc 
(e.g., Rasmussen et al.\,2003).
It therefore remains unknown whether the coronal gas really extends to the virial 
radius of the Milky Way ($R_{\rm vir}\sim 260$ kpc; e.g., Tepper-Garc{\'{\i}}a et al.\,2015) 
and how it connects to gas gravitationally bound to the Local Group. From
the UV observations of O\,{\sc vi} in the thick disk of the Milky Way 
(Widmann et al.\,1998; Savage et al.\,2003; Wakker et al.\,2003; Savage \& Wakker 2009)
indeed follows that there are large amounts of extra-planar warm-hot gas at vertical heights 
$z<5$ kpc.

Indirect evidence for the widespread presence of hot gas in the Milky Way halo comes
from the many UV absorption-line detections of {\it high-velocity} O\,{\sc vi} 
(Sembach et al.\,2003; Wakker et al.\,2003), which has a sky covering fraction as 
large as $f_{\rm c}=0.60-0.85$. O\,{\sc vi} arises in warm-hot gas at $T\sim 3\times 10^5$ K
and is believed to
trace the interface regions between the hot coronal gas and cooler halo clouds
embedded therein (i.e., the neutral and ionized HVCs; see above). Isolated,
strong O\,{\sc vi} absorption at high velocities is also seen in the direction of the Local Group
barycenter and, in particular, towards the quasar Mrk\,509 in HVC Complex GCN 
(Fig.\,5), where enhanced X-ray emission and high-velocity 
Si\,{\sc iii} absorption is observed (Collins, Shull \& Giroux 2005; see above). The coincidence of 
UV absorption and X-ray emission further indicates that there is an
excess of hot gas in this direction that possibly is related to highly-ionized 
intragroup gas near the Local Group barycenter, towards which the Milky Way is 
moving (see Sect.\,2.2). 

To estimate the baryon content of the Milky Way's coronal gas, its radial density 
profile needs to be constrained. Due to on-going accretion of gas and stars from 
satellite galaxies and the IGM, substantial deviations from a simple hydrostatic
density distribution are likely.
Even for a hot halo that is not perfectly hydrostatic, however, 
the average gas density in the coronal gas 
is expected to decrease for increasing distances to the disk. From X-ray 
absorption, spectra Bregman \& Lloyd-Davies (2007) estimate a gas density of
$n_{\rm H}\approx 8\times 10^{-4}$ cm$^{-3}$ for the inner halo ($d<20$ kpc). From
pulsar dispersion measures instead follows that the average coronal 
gas density must be smaller, $n_{\rm H}< 8\times 10^{-4}$ cm$^{-3}$, in
line with studies that estimate $n_{\rm H}$ indirectly from considering
the interaction between the cool HVCs and the ambient hot medium 
($n_{\rm H}\approx 2\times 10^{-4}$ cm$^{-3}$; Grcevich \& Putman 2009;
Peek et al.\,2007; Tepper-Garcia et al.\,2015). 
Because of the unknown extent and the unknown gas properties
at the virial radius, the total mass of the Milky Way's hot coronal gas is 
very uncertain. For $d<250$ kpc the total mass is estimated to be
$M_{\rm Corona}\approx 10^{10}-10^{11} M_{\odot}$ (Anderson \& Bregman 2010; Yao et al.\,2008;
Gupta et al.\,2012; Miller \& Bregman 2013, 2015; Fang, T., Bullock \& Boylan-Kolchin 2013;
Salem et al.\,2015),
in line with the idea, the Milky Way's hot Corona represents a 
huge baryon reservoir. However, the hot coronal gas must cool and condense
into streams of denser gas to be able to sink to the Milky Way disk,
i.e., it must transit through the diffuse ionized and/or neutral phase to 
contribute to the gas-accretion rate. To understand the details of this
important phase transition, hydrodynamical simulations are required, which
will be discussed in Sect.\,3.


\begin{figure}[t!]
\centering
\includegraphics[width=12cm]{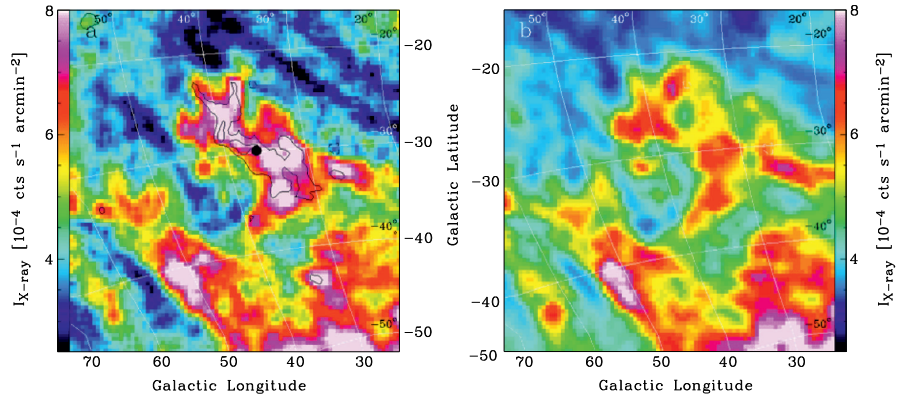}
\caption{
{\it Left panel:} X-ray emission map of hot gas in the direction of
HVC complex GCN, based on ROSAT $0.25$ keV soft-X ray background (SXRB)
data (see Kerp et al.\,1999 for details). The position of the background
quasar Mrk\,509 is indicated with the black dot. In this direction,
hot halo gas has been detected in high-velocity O\,{\sc vi} absorption using
FUV data (Sembach et al.\,2003; see also Winkel et al.\,2011).
{\it Right panel:} model of the SXRB emission in the same direction,
based on 21 cm H\,{\sc i} data from foreground gas (causing photoelectric
absorption of the background X-ray photons) and assuming a
homogenous distribution of the X-ray background (Kerp et al.\,1999).
Both maps indicate the presence of hot coronal gas in the Galactic halo
in this area of the sky. Maps kindly provided by J\"urgen Kerp.
}
\label{fig:5}       
\end{figure}


\subsection{Gas-accretion rates from observations}
\label{subsec:5}

The observations presented in the previous sections suggest that HVCs (and IVCs) 
represent coherent entities of multi-phase gas that move within the Milky Way halo
(Lehner et al.\,2012; R17). Independent of their individual
origin (tidal interactions with satellite galaxies, infall from the IGM or
intragroup medium, condensations from the hot coronal gas, the back-flow
of material expelled previously from the disk), these streams of gas 
trigger the Galaxy's present-day gas-accretion rate as defined in 
equation (1) and we will discuss their contribution to 
d$M_{\rm gas,halo}$/d$t$ in the following.

While the 3D distribution of neutral and diffuse ionized gas around the Galaxy
(i.e., $M_{\rm gas,halo}$, $d$) is constrained by observations, the space 
{\it motion} of the gas is not, as only the radial component of the velocity
can be observed. In lack of further information, the mean infall 
velocity often is assumed to be constant for all halo clouds,
although it is likely that $v_{\rm infall}$ spans a large range for the IVCs and 
HVCs and systematically depends on $d$ (see Sect.\,1).
For the MS, recent studies assume $\langle v_{\rm infall}\rangle =
100$ km\,s$^{-1}$ (e.g., Fox et al.\,2014), a value
that also has been used for other HVCs (e.g., Complex C; Wakker et al.\,1999, 2008).
A more complex model for $v_{\rm infall}$ is presented by Putman, Peek \& Joung (2012), 
where they try to separate for the HVC population the azimuthal velocity component 
from the accretion velocity in the Galactic center (GC) direction and derive 
$\langle v_{\rm infall,GC} \rangle\approx -50$ km\,s$^{-1}$.
This value is consistent with the {\it mean} radial velocity of HVCs of 
$\langle v_{\rm rad} \rangle\approx -50$ km\,s$^{-1}$, but 
the most distant halo structures, such as the MS, might have
somewhat larger accretion velocities (Mathewson et al.\,1974).

Because of its large mass, the MS dominates by far the total present-day gas
accretion rate of the Milky Way. From their UV absorption-line survey
Fox et al.\,(2014) derive a mass
inflow rate of neutral and diffuse ionized gas from the Stream of
d$M_{\rm MS}/$d$t \approx 2\,M_{\odot}\,$yr$^{-1}$ for a fixed
distance of the MS of $d=55$ kpc and $\langle v_{\rm infall}\rangle =100$ km\,s$^{-1}$.
At this distance and infall velocity, the gas (or better said, a
fraction $\eta$ of it; see equation (2)) would
reach the disk in $\sim 540$ Myr. If, instead, the distance of the Stream was
$d=100$ kpc, then it would take $\sim 1$ Gyr for the gas to reach the disk and
the accretion rate would be higher by a factor of $\sim 2$, because the estimate
of $M_{\rm MS}$ depends on its assumed distance.
Taking $d=55$ kpc as a conservative lower limit and adding the gas mass
associated with the Magellanic Clouds and the Magellanic Bridge (with
all these components together forming the {\it Magellanic System}),
the total accretion rate from all these components sums up to a value of
d$M_{\rm MSys}/$d$t \geq 3.7\,M_{\odot}\,$yr$^{-1}$; Fox et al.\,2014)

Based on the observed properties discussed above, the contribution of the 
other {\it individual}, 21 cm-selected HVC Complexes at $d<20$ kpc to the mass-inflow rate 
(including both neutral and ionized gas) is expected to be rather small
(e.g., Complex C: $0.1-0.2 \,M_{\odot}\,$yr$^{-1}$; Complex A: $0.05 \,M_{\odot}\,$yr$^{-1}$;
Cohen Stream: $0.01\,M_{\odot}\,$yr$^{-1}$; see Wakker et al.\,1999, 2007, 2008; 
Thom et al.\,2006, 2008; Putman, Peek \& Joung 2012; R17). 
Putman, Peek \& Joung (2012) derive a maximum accretion rate of $0.4\,M_{\odot}\,$yr$^{-1}$
for all HVCs {\it except} the MS.
If we consider only the neutral
gas mass in the Galactic HVC population, the total H\,{\sc i} gas accretion
rate from all 21 cm HVCs (including the MS) comes out to d$M_{\rm HI}/$d$t =0.7\,M_{\odot}\,$yr$^{-1}$ 
(Richter 2012), again assuming that the MS is at $d=55$ kpc.

The general distribution of UV-absorbing gas in the halo (and its total mass)
yet implies, that the {\it ionized} component (independent of whether it is associated 
with H\,{\sc i} emission or not) dominates the gas accretion not only for the MS, but 
also for the nearby HVCs at $d\leq20$ kpc.
From their absorption-line survey towards halo stars and extragalactic background sources,
Lehner \& Howk (2011) determine an accretion rate of predominantly ionized high-velocity gas 
at $d<15$ kpc of $0.45-1.40\,M_{\odot}\,$yr$^{-1}$.
Putting it all together, R17 estimate from their UV absorption-line
survey of 265 sightlines 
a total gas-accretion rate of neutral and ionized high-velocity gas in the
halo (including gas from the Magellanic System) of 
d$M_{\rm HVC}/$d$t \geq 5\,M_{\odot}\,$yr$^{-1}$.
This limit is higher by a factor of $>2$ than the {\it current} star-formation 
rate of the Milky Way ($\sim 0.7-2.3 M_{\odot}$\,yr$^{-1}$; e.g., Robitaille \& Whitney 2010;
Chomiuk \& Povich 2011).

The contribution of the 21 cm IVCs and the DIG in the disk-halo 
interface (DHI) at $z$-heights $<2.5$ kpc to the Milky Way's gas-accretion rate 
is significantly smaller than that of the neutral HVCs. Considering the IVC distances 
and neutral-gas masses discussed in Sect.\,2.1.1, the neutral gas-accretion rate from IVCs is 
only $\sim 0.01-0.05 M_{\odot}$\,yr$^{-1}$. A much higher gas-accretion rate 
can be determined considering the ionized gas reservoir in the DIG (Sect.\,2.2). 
In principle, the overall gas flow in the DIG of Milky-Way type is expected to be 
strongly influenced by the various feedback processes from the disk (e.g., radiative
and mechanical feedback from supernovae, stellar winds, and AGN; e.g.,
Bland-Hawthorn \& Maloney 2002; MacLow \& Klessen 2004; Springel \& Hernquist 2005;
Marasco, Marinacci \& Fraternali 2013), but the exact role of these processes 
in the gas-circulation cycle of the Milky Way's DHI is uncertain. Assuming that at least
half of the diffuse ionized gas in the disk-halo interface is currently being accreted 
(the rest being related to outflowing gas), the observational constraints imply 
d$M_{\rm DHI}/$d$t = 1-2\,M_{\odot}\,$yr$^{-1}$ for $v_{\rm infall}\leq 20$ km\,s$^{-1}$,
thus in line with the current star-formation rate (see also Fraternali et al.\,2013).
A net-infall of ionized gas is in line with the observed H$\alpha$ kinematics
(Haffner et al.\,2003). 

Some interesting conclusions can possibly be drawn from these numbers.
Obviously, the amount of large-scale {\it neutral} gas in the disk-halo interface 
is much smaller than the amount of large-scale neutral gas at larger distances; it is also much
smaller than the amount required to keep up the current star-formation rate in the disk. 
Infalling neutral gas structures thus might be disrupted and ionized when
entering the disk-halo interface, where it might re-cool and condense again
before it enters the disk. This re-processing of infalling gas in the
disk-halo interface is sometimes referred to as ``quiet accretion''.  
The galactic-fountain flow is believed to play a crucial role in the
the gas cooling and condensation processes (Marinacci et al.\,2010; 
Armilotta, Fraternali \& Marinacci 2016). The infalling gas thus might 
enter the disk in the form of low-velocity, mildly-ionized gas clumps
or tiny 21 cm drops (Lockman 2002; Begum et al.\,2010; Ford, Lockman
\& McClure-Griffiths; see Sect.\,2.1),
thus in a form that is difficult to identify observationally.

Another possible reason for the apparent discrepancy between the neutral gas budget
in the DHI and that at larger distances might be the existence of neutral DHI gas 
that is ``hidden'' to us: clouds that have low radial velocities 
(LVCs; see Sect.\,1.1), similar to those in the disk, 
but that reside in the (lower) Galactic halo (e.g., Zheng et al.\,2015; Peek et al.\,2009).
Finally, also gas from the outer disk might contribute to the fueling of star formation
in the inner regions of the Milky Way disk through a {\it radial} inflow of gas (e.g.,
Elson et al.\,2011; Sellwood \& Binney 2002; see also Putman, Peek \& Joung 2012
and references therein).

Obviously, observations alone cannot provide a full insight into the complex processes
that govern the past, present, and future gas-accretion rate of the Milky Way.
Hydrodynamical simulations represent an important toolkit to further study 
the dynamics and physical properties of gas falling toward Milky-Way type galaxies
and to pinpoint the overall mass-inflow rate to the disk.
The most relevant of these aspects will be discussed in the following section.


\section{Simulations of Milky Way gas accretion}
\label{subsec:3}

In this section, we briefly discuss results from theoretical studies of accretion 
processes of Milky Way-type galaxies based on hydrodynamical simulations. Although theses 
studies will also be discussed in a more general context in other chapters in the 
second part of this book, they are particularly important for our understanding of gas 
accretion in the Milky Way and thus need to be considered in this review.

\subsection{Hydrodynamical simulations of gas infall}
\label{subsec:1}

Whatever the initial conditions of an infalling gas structure at a given
distance to the Galaxy might be, the gas will interact with the ambient
hot coronal medium that fills the galaxy's dark-matter potential.
Modeling these interaction processes are of fundamental importance to
understand origin and fate of accreted material and to explain the 
observations discussed in the previous sections.

As mentioned earlier, one plausible scenario for gas accretion in Milky-Way size
gaseous halos with $M_ {\rm halo}\sim 10^{12}\,M_{\odot}$ and 
$R_{\rm vir}\sim 250$ kpc is the
cooling and fragmentation of hot ($T\sim 10^6$ K) halo gas
that falls towards the Galactic disk in the form of overdense clouds 
(Maller \& Bullock 2004). Support for this scenario comes from early SPH
simulations (Kaufmann et al.\,2006, 2009; Sommer-Larsen 2006).
In these simulations, cool pockets of gas condense out from the hot coronal 
gas from thermal instabilities, mimicing the properties of the 
Galactic 21 cm HVCs that have $d<20$ kpc (Peek, Putman \& Sommer-Larsen 2008).
More recent studies indicate, however, that these early studies may draw a too
simplistic picture of the HVC condensation process.
One problem lies in the buoyancy of thermally unstable gas,
which is expected to disrupt condensing seed structures before they can
cool efficiently (Burkert \& Lin 2000). Thus, linear isobaric perturbations
in a homogeneous, hot coronal gas are expected to be inefficient 
to develop cool gas patches in the halo, unless
the entropy gradient is very small. An alternative scenario is offered
by Joung, Bryan \& Putman (2012), who study the condensation process in 
Galactic coronal gas from non-linear perturbations that might be generated 
from either infalling intergalactic gas (e.g., Kere{\v s} \& Hernquist 2009;
Kere{\v s} et al.\,2009; Brooks et al.\,2009) or from the gaseous leftovers of 
satellite accretion (Bland-Hawthorn et al.\,2007; Grcevich \& Putman 2009;
Nichols \& Bland-Hawthorn 2011;
see below). Joung, Bryan \& Putman (2012) find from their high-resolution 
adaptive mesh refinement (AMR) hydrodynamical simulations that the efficiency for 
condensing out cool gas patches from non-linear perturbations depends critically
on the ratio between the cooling time and the acceleration time (to reach the 
sound speed) in the gas. If cooling is efficient (such as in clumpy gas and/or in gas
with a high-metallicity), cool patches can condense out in a Milky Way-type halo 
before being disrupted and may show up as 21 cm HVCs. The mixing of 
outflowing, metal-enriched fountain material and infalling coronal 
material possibly represents a key process that determines the net accretion
through the disk-halo interface, as it regulates the cooling 
efficiency of the gas (Armilotta, Fraternali \& Marinacci 2016; Marinacci et al.\,2010).

Cosmologically ``cold'' ($T<T_{\rm vir}$) streams of intergalactic gas are 
expected to feed the 
halos of Milky Way-type galaxies (e.g., van\,de Voort et al.\,2011) and such streams 
may penetrate deep into the hot halo before they are being disrupted.
From their AMR simulations,
Fernandez, Joung \& Putman (2012) find that cold streams from the IGM may 
continuously bring up to $\sim 10^8 M_{\odot}$ of H\,{\sc i} into the Milky Way 
halo and this amount of neutral gas appears to be fairly constant over 
time (at least for the last 5 Gyr, or $z=0-0.45$). The H\,{\sc i} gas 
accretion rate for a Milky Way-type galaxy in these simulations comes
out to $0.2\,M_{\odot}$\,yr$^{-1}$, thus very similar
to what has been derived for the Galactic HVC population at $d<20$ kpc, i.e.,
{\it without} the contribution of the MS (see Sect.\,2.4).

Next to the cloud-condensation scheme and the cold streams, 
the interaction of Milky-Way type galaxies 
with dwarf satellites can transport large amounts of relatively cool gas in the 
halos of MW-type galaxies {\it directly}.
From their study of a Milky-Way type galaxy and its satellite-galaxy 
environment Fernandez, Joung \& Putman (2012) derive
a median mass-loss rate of H\,{\sc i} of $\sim 3 \times 10^{-3}\,M_{\odot}$\,yr$^{-1}$,
suggesting that satellites add $\sim 3\times 10^6\,M_{\odot}$ of neutral gas to 
the host galaxy within a Gyr. Based on their overall results, this H\,{\sc i} mass flow 
represents a significant albeit not dominating contribution to the neutral gas accretion
rate of the host galaxy. In the light of these results, the observed neutral gas
supply from Magellanic Stream for the Milky Way ($M_{\rm HI,MS}\sim 2.5\times 10^8\,M_{\odot}$
within $0.5-1.0$ Gyr); see Sect.\,2.1) is huge, underlining that the MS represents a 
rather extreme (and atypical) example for gas accretion by satellite interactions.

As suggested by Peek (2009), the specific angular momentum of the gas might be a key
parameter to discriminate between the various scenarios for the origin of the HVCs. 
Gas that is accreted from dwarf satellites by either ram-pressure
stripping or tidal interactions enters the halo with a large initial
angular momentum ($L \sim 3\times 10^4$ km\,s$^{-1}$\,kpc), while the 
angular momentum of the cooling halo clouds is one order of magnitude
less (see, e.g., Peek et al.\,2008; Kaufmann et al.\,2009).
Thus, gas accretion from mergers with Milky Way satellites takes
place predominantly in the outer regions ($R>15$ kpc) of the Galactic disk.
In fact, observations indicate that part of the MS (i.e., the Leading Arm) 
might already be close to the outer disk of the Milky Way, where it possibly 
already interacts with the underlying interstellar gas (McClure-Griffiths et al.\,2008;
Casetti-Dinescu et al.\,2014), thus providing support for this scenario.
Following Peek (2009), the HVCs at small galactocentric distances ($R<15$ kpc), 
in contrast, are more likely to be produced by the condensed halo clouds and 
it is this process that appears to dominate the feeding of the inner-disk regions
where most of the star formation takes place.

Independently of the origin of the infalling gas, how much of it makes it into the disk?
To answer this crucial question for the Milky Way and other galaxies of 
similar mass and type, the initial conditions for the gas infall 
(e.g., infall velocities, individual cloud masses, density profile of the
ambient coronal gas) turn out to be particularly important, as we will
discuss in the following.

At $T=10^6$ K in the coronal gas, the infall velocity of the HVCs 
is close to the sound speed in ambient hot medium ($v_{\rm s}\sim 150$ km\,s$^{-1}$),
so that infalling material moves either in the subsonic, or transonic, or 
supersonic regime, leading to different ablation scenarios of infalling gas
structures (Kwak et al.\,2011). Using a grid-based hydrodynamical 3D code,
Heitsch \& Putman (2009) simulated the fate of H\,{\sc i} clouds moving through
the hot coronal gas to explore the characteristic morphologies of the infalling structures,
such as head-tail structures, infall velocities, disruption path lengths, and timescales.
In Fig.\,6 we highlight some of their results.
From their study it follows that H\,{\sc i} clouds with relatively low initial 
masses of $M_{\rm cloud}<10^{4.5} M_{\odot}$ lose basically all their neutral
gas content when infalling from $d=10-12$ kpc. Thus, wherever such cloudlets might 
be formed in the halo, i.e., either from condensing out of the coronal gas or 
from dissolving infalling tidal or cosmological gas streams at much larger distances, 
their gas content will not reach the disk in {\it neutral} form. 
However, if the density contrast between the break-up material and the surrounding 
medium is large enough, the cloud remnants might reach
the disk as warm ionized material (e.g., Shull et al.\,2009; Heitsch \& Putman
2009; Bland-Hawthorn 2009; Joung et al.\,2012) or will otherwise fuel the Galactic Corona.


\begin{figure}[t!]
\centering
\includegraphics[width=11cm]{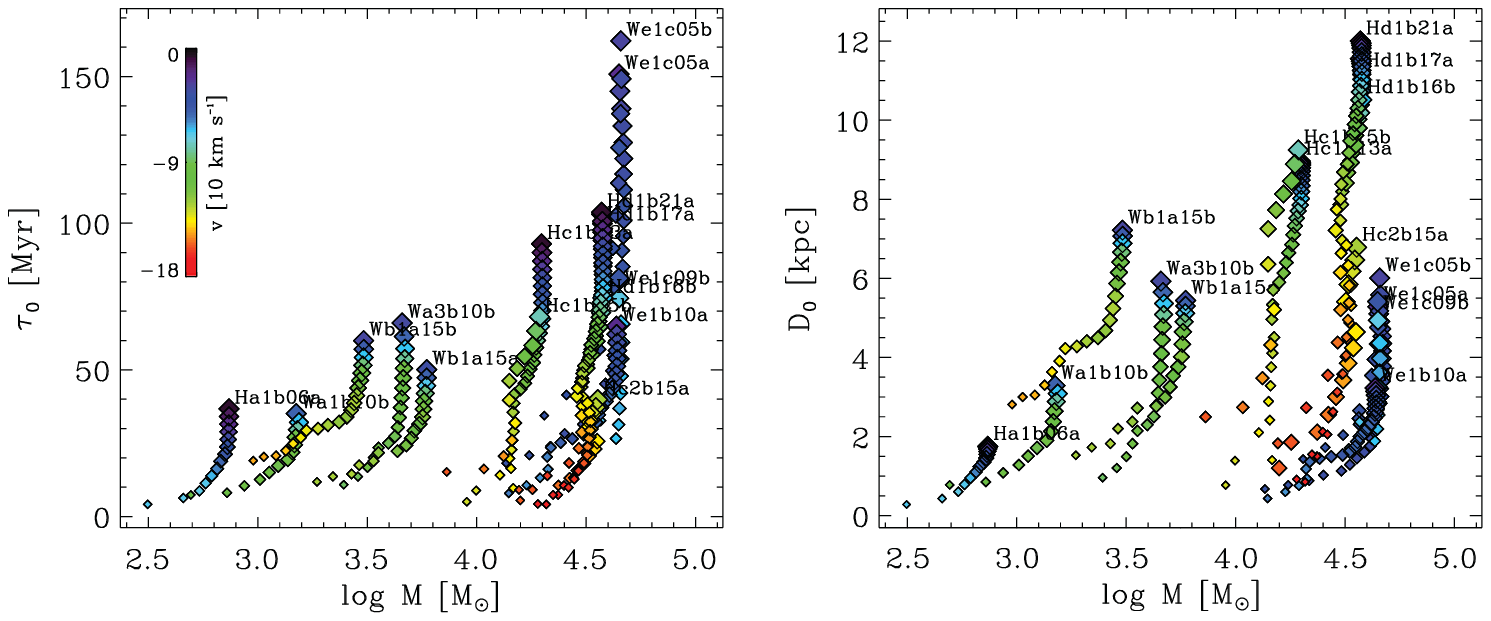}
\caption{
Time sequences of cloud survival-times (left panel) and disruption
length-scale (right panel) 
of neutral halo clouds of different mass that move through hot coronal
gas in a Milky-Way type galaxy (Heitsch \& Putman 2009). In these
models, H\,{\sc i} clouds with masses $<10^{4.5}M_{\odot}$ do not
survive their passage through the halo when infalling from
$d=10-12$ kpc. Figure adopted from Heitsch \& Putman (2009).
}
\label{fig:6}       
\end{figure}


At the high-mass end, Kwak et al.\,(2011) predict that
for cloud masses $M_{\rm cloud}>10^5 M_{\odot}$ up to 70 percent of the
H\,{\sc i} mass remains cool at $T<10^4$, even after a
possible break-up of the initial infalling cloud.
Such massive structures may even survive the trip from larger distances,
as they can move several hundred Myr through the hot halo before
being destroyed completely. A similar conclusion was drawn from
Joung, Bryan \& Putman (2012) from their high-resolution AMR simulations
(see above). This result is relevant for 
the Magellanic Stream at $d=50-100$ kpc. Although observations suggest
that the MS is further braking up into smaller gas clumps as it moves
towards the Milky Way disk (e.g., Stanimirovic et al.\,2002, 2008; 
Westmeier \& Koribalski 2008; see also Tepper-Garcia et al.\,2015), 
some dense cores with masses $M_{\rm cloud}>10^5 M_{\odot}$ may survive and 
these clumps could enter the disk in the form of neutral gas clouds.

Note that a certain fraction of the halo clouds that condense out of the 
coronal gas may also reach the disk {\it before}
the gas cools efficiently, i.e., in the form of warm, ionized 
gas that never becomes neutral and visible in 21 cm emission.
Those structures possibly explain some of the isolated high-velocity UV absorber
that have no 21 cm counterparts (Collins, Shull \& Giroux 2003; 
Collins et al.\,2009; Shull et al.\,2009; 
Richter et al.\,2009; Fraternali et al.\,2013;
Marasco, Marinacci \& Fraternali 2013; R17).

Even with the most advanced hydrodynamical simulations at hand,
the exact role of many of the involved physical processes in the 
time evolution of accreted gas remains unclear.
For instance, thermal conduction can suppress shear instabilities 
and thus stabilize clouds from evaporation by smoothing
out steep density/temperature gradients between the cool
infalling gas and the hot ambient medium (e.g., Vieser \& Hensler 2007).
However, such steep gradients appear to be unimportant for
the cool/warm gas clouds in the Milky Way halo (e.g., Kwak et al.\,2011)
and also the diffusion by thermal conduction possibly is 
small compared to turbulent diffusion processes.
Not all of the possibly relevant processes can be included simultaneously 
in the simulations.
In particular, the role of magnetic fields (and resulting
magnetohydrodynamical effects) have been mostly ignored so far.
Also, the spatial (or mass) resolution of many of the simulations discussed above
still is  very limited. Since early SPH simulation codes, for instance,
had problems in resolving Kelvin-Helmholtz instabilities,
cool circumgalactic gas clumps are artificially stabilized
by SPH particle effects (Agertz et al.\,2007), probably leading to 
misleading results.

In summary, hydrodynamical simulations predict that the life-time,
the morphology, and the mass distribution of infalling gas clouds in the Milky Way
halo depend strongly on the local boundary conditions under which the
gas is generated and moving through the hot halo.
Whether or not an infalling cool/warm gas cloud reaches the Milky Way disk in a region
where its supplements star formation depends on its initial distance to the disk,
its mass and density, its infall velocity, its angular momentum, and other parameters.

\subsection{Cosmological hydrodynamical simulations}
\label{sec:3}

Given the complexity of the physics of gas accretion outlined above and the strong
dependence of the gas-accretion rate on the local boundary conditions,
additional large-scale simulations are desired that also consider a realistic 
{\it cosmological} environment of the Milky Way. In particular, the role of
a second nearby spiral galaxy (M31), the gas motions within the
super-ordinate galaxy group environment (Local Group), as well as the 
streaming of intergalactic gas within the local cosmic web that connects the
Local Group with its surrounding large-scale structure (e.g., the Virgo cluster)
should be studied in such a context.

In a recent paper, N14 have studied the large-scale
distribution and overall physical properties of gas in the Local Group and around 
Milky Way and M31 based on simulation data from the Constrained Local 
UniversE Simulations (CLUES) project (www.clues-project.org).
In their study, the authors separate the circumgalactic and intragroup 
gas into three different phases: neutral gas, cold/warm ionized gas with
$T<10^5$ K, and hot gas with $T\geq 10^5$ K, similar as done here.
The total neutral gas mass in the simulated Milky Way at galactocentric
distances $d<50$ kpc comes out to $M_{\rm HI}\approx 3\times 10^8 M_{\odot}$,
thus in excellent agreement with the observations (Sect.\,2.1). The total
mass of the cold/warm ionized gas component is as large as
$M_{\rm HII}\approx 3\times 10^{10}\,M_{\odot}$ for the entire halo
out to the virial radius, but reduces to 
$M_{\rm HII}\approx 2\times 10^8 M_{\odot}$ for $d\leq 10$ kpc, where
the bulk of UV absorption of warm-ionized gas is observed in the
Milky Way (Lehner \& Howk 2011; Sect.\,2.2). The mass of the Milky Way's 
hot coronal gas in the simulation is 
$M_{\rm Corona}\approx 4\times 10^{10}\,M_{\odot}$ for $d\leq R_{\rm vir}$
and $\approx 10^{10}\,M_{\odot}$ for $d\leq 100$ kpc, the latter value being 
consistent with the estimates from X-ray observations (Sect.\,2.3).
The Milky Way's neutral gas-accretion rate from gas at $d\leq 50$ kpc 
is estimated as $M_{\rm HI}\approx 0.3\,M_{\odot}$\,yr$^{-1}$, which is 
only $\sim 40$ percent of the value derived from the 21 cm observations
(Richter 2012), but more in line with what is expected for the neutral
HVCs without the Magellanic Stream (Sect.\,2.4). For larger distances to the disk,
the neutral gas-accretion rate quickly falls below $10^{-2}\,M_{\odot}$\,yr$^{-1}$
in the simulations (N14; their Fig.\,14). The
accretion rate of cold/warm gas instead is fairly independent of the distance for
$d>15$ kpc at a level of $M_{\rm HI}\approx 5\,M_{\odot}$\,yr$^{-1}$,
thus in line with the estimate from the UV 
observations (Fox et al.\,2014; R17).
Finally, the simulations imply that only for very large distances 
$d>100$ kpc the accretion rate of hot ($T>10^6$ K) gas dominates
the mass inflow of gas for the Milky Way. 

The influence of the Local Group environment in the simulations
is reflected particularly in the anisotropic distribution of gas near the 
viral radius of the MW, because the gas follows the large-scale matter 
distribution in the elongated cosmological filament that forms the Local 
Group (N14). To visualize this, we show in Fig.\,7 
the gas distribution and gas 
kinematics around the simulated Milky Way and M31 galaxies from the
CLUES simulations.
The two galaxies move towards the LG barycenter while the ambient gas 
is circulating around MW and M31 within the elongated filament 
in a complex pattern of infall and outflow channels.
There is a significant gas excess between the two galaxies, 
as compared to any other direction, resulting from the overlap of
their gaseous halos. Because of the Milky Way's flow towards 
LG barycenter, a velocity dipole pattern for high-ion absorption 
from LG gas/M31 halo gas is expected from a perspective within the 
Milky Way (R17).
Such a dipole pattern is indeed observed in UV absorption in the 
direction of M31 and its antipode on the sky (Sect.\,2.2). 
This possibly implies that warm/hot LG gas/M31 halo gas 
is pushed into the Milky Way halo due to the large-scale motion of 
both galaxies in their group environment. If true, this effect might 
cause a major boost in the Milky Way's future gas-accretion rate.


\begin{figure}[t!]
\includegraphics[width=12cm]{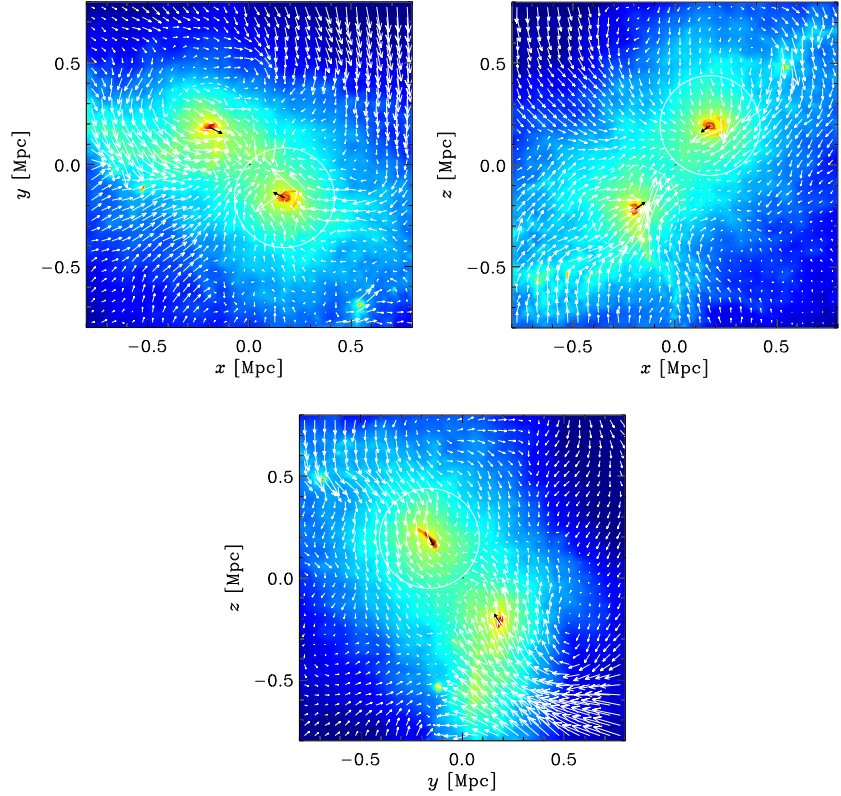}
\caption{
Projected distribution and bulk motions of gas in and around the 
Milky Way (dashed circle, indicating the virial radius of the MW) 
and M31 (solid circle, indicating the virial radius of the M31) 
based on constrained cosmological simulations from the 
CLUES project (N14).
Coordinates and velocities of the gas are given with respect
to the Local Group barycenter in the $x/y$, $x/z$, and $y/z$ 
planes, respectively.
The white arrows show the velocity field of the gas, with the longest arrows
representing a velocity of $130$ km\,s$^{-1}$.
The black arrows indicate the velocities of the MW and M31 galaxies.
Their absolute space velocities are $67$ and $76$ km\,s$^{-1}$,
respectively. Maps kindly provided by Sebasti\'an Nuza.
}
\label{fig:7}       
\end{figure}


\subsection{Comparison with observations}
\label{sec:2}

For a better understanding of the gas-accretion processes in the Milky 
Way, the comparison between observational data and predictions from
simulations are essential. Next to the gas masses in the 
individual phases and the accretion rates (see above), 
the spatial distribution of the various gas phases in the simulations 
and their kinematics can be compared 
(in a statistical sense) with the observational constraints from 21 cm
data and UV absorption spectra.

Combining 21 cm data from the Milky Way and M31, Richter (2012) predicted 
that the volume-filling factor of neutral gas in the halo of Milky-Way/M31
type galaxies declines exponentially with radius, leading to an
exponential decline of the observed (projected) H\,{\sc i} covering fraction.
The study suggests that the covering fraction $f_{\rm c}$(H\,{\sc i}) drops 
below $0.05$ for $d>50$ kpc. A similar trend indeed has also been found 
in the recent CGM simulations of Milky-Way type galaxies (Fernandez, Joung \& 
Putman 2012; N14), indicating that basically all neutral 
gas in MW-type galaxies is concentrated in the inner halo region. This
conclusion is supported by the observed cosmological cross-section of 
neutral gas around low-redshift galaxies (Zwaan et al.\,2005; Richter et al.\,2011).

Also the apparent interaction between the MS and the surrounding coronal 
gas as well as the H$\alpha$ emission from the Stream have been investigated in simulations 
to reproduce the observational results. Bland-Hawthorn et al.\,(2007) and 
Tepper-Garcia et al.\,(2015) modeled the H$\alpha$ emission from the MS 
based on a shock-cascade model. Tepper-Garcia et al.\,(2015) conclude
that the H$\alpha$ emission from the Stream can only be reproduced if
the density of the ambient medium is $n_{\rm H}=2-4\times 10^{-4}$ 
cm$^{-3}$, indicating that  H$\alpha$ emitting regions in the MS must be within
$d\leq 75$ kpc from the Galactic center.

Next to these examples, there are several other studies that have addressed
these and other aspects by comparing observational results with simulations.
Describing all of these unfortunately is beyond the scope of this review.
Clearly, with future, more detailed simulations and additional
constraints from multi-wavelength observations the systematic combination
of simulations and observations will provide crucial new insights into
the properties of the Milky Way's CGM.

\section{Concluding remarks}
\label{sec:5}

As for many other aspects of galaxy evolution, the Milky Way and its gaseous 
environment represent an excellent laboratory to study the details of 
gas-accretion processes of $L^{\star}$ galaxies in the local Universe.
Although it remains a challenging task to reconstruct the 3D distribution and 
galactocentric kinematics of the Milky Way's circumgalactic medium from an
internal vantage point in the rotating disk, the combination of 
multi-wavelength observations of the gas in all its phases, the measurement and 
modeling of the stellar composition of the Milky Way disk and its
star-formation rate and history, the numerical modeling of the
hydrodynamic processes that shape the properties of the Galaxy's CGM, and
the deep observations of the Milky Way and its satellite galaxies in a
cosmological context together provide a particularly rich database 
that cannot be achieved for any other galaxy in the Universe.

The above discussed observations and simulations imply that the combination 
of gas infall, outflows, and mergers generates a multi-phase gaseous halo that is 
characterized by a highly complex spatial distribution of gas structures of 
different age and origin.
The cycle of processes that is believed to lead to the continuous feeding of the Milky Way disk 
with fresh gas to supplement subsequent star-formation therein can be summarized as follows.

Cold and warm gas from the intergalactic medium and from satellite galaxies enters
the Milky Way halo at its virial radius and is then processed by the ambient hot coronal gas.
Fragments of the originally infalling gas may reach the disk in the form of warm or cold
gas streams, while the remaining gas fraction is being incorporated into the hot Galactic 
Corona. In this way, the Corona is continuously fed with fresh gas from outside, while
it is further stirred up and heated by gas outflowing from the star-forming disk and 
(eventually) from the Galactic center region (i.e., it is influenced by feedback processes).
From the hot Corona, warm ionized and/or warm neutral gas patches may condense out 
through cooling processes, and these structures will sink down to the disk through
the disk-halo interface, further contributing to the overall accretion rate of the
Milky Way disk. 

The accretion of cold and warm gas from the Magellanic Stream is a direct result of 
the interaction between the Milky Way and its population of satellite galaxies 
(here: the Magellanic Clouds). The MS adds more than 1 billion solar masses 
of gaseous material to the Milky Way halo, material that either directly or indirectly
feeds the MW disk to supplement star formation therein.
Thus, there is sufficient cold and warm gaseous material present in the outer
halo to maintain the Milky Way's star-formation rate {\it in the far future} at its
{\it current} level, although it remains unclear, how much of the material from the
MS will finally end up in the disk (and at what time scale).
The amount of gas that is currently being accreted through
the disk-halo interface, and that will determine the star-formation rate {\it in the near 
future}, remains uncertain, however.

Our understanding of the gas-accretion processes in the Milky Way is far from 
being complete. On the observational side, additional constraints on distances and 3D
velocities of the HVCs, the role of low-velocity halo gas, and the mass and spatial extent 
of the hot coronal gas based on multi-wavelength observations are highly desired
on the long way towards a complete census of the Milky Way's circumgalactic gas.
On the theoretical side, more advanced hydrodynamical simulations of the Milky Way's
gaseous halo, that include all relevant physical processes in a realistic cosmological 
environment, will be of great importance to study the dynamics of gas flows around 
the Galaxy.
Finally, a systematic comparison between gas-accretion processes in the Milky Way and 
similar processes in other low- and high-redshift galaxies in the general context
of galaxy evolution (with particular focus on environmental issues, feedback effects, 
and other important aspects) will provide crucial information that will help to better
understand the cycling of gas on galactic and super-galactic scales.
Many of these aspects will also be discussed in the following chapters.


\begin{acknowledgement}
The author would like to thank Andy Fox, Matt Haffner, Fabian Heitsch, J\"urgen Kerp, 
\& Sebasti\'an Nuza for providing helpful comments and supplementary material for 
the figures.
\end{acknowledgement}




\end{document}